\documentclass[tightenlines,aps,prc,showpacs,amsmath,showkeys,
superscriptaddress]{revtex4}

\usepackage[dvips]{graphicx}

\begin{document}

\title{Overview of pion-nucleus interaction at low energies}
\author{E.~Friedman}
\email{elifried@vms.huji.ac.il}
\affiliation{Racah Institute of Physics, The Hebrew University,
Jerusalem 91904, Israel}

\date{\today}

\begin{abstract}
Predictions of the existence of well-defined deeply bound pionic 
atom states for heavy nuclei 
and the eventual observation of such states by the
$(d,^3He)$ reaction have revived interest in the pion-nucleus interaction
at threshold and in its relation to the corresponding pion-nucleon interaction.
Explanation of the `anomalous' $s$-wave repulsion in terms of partial
restoration of chiral symmetry  and/or in terms of energy-dependence 
effects have been
tested in {\it global} fits to pionic atom data and in a recent
{\it dedicated} elastic scattering experiment. The role of neutron density
distributions in this context 
is discussed in detail for the first time.

\end{abstract}

\maketitle

\section{Motivation and Background}
\label{sec:mot}
\begin{figure}
\includegraphics[height=8cm,width=10cm]{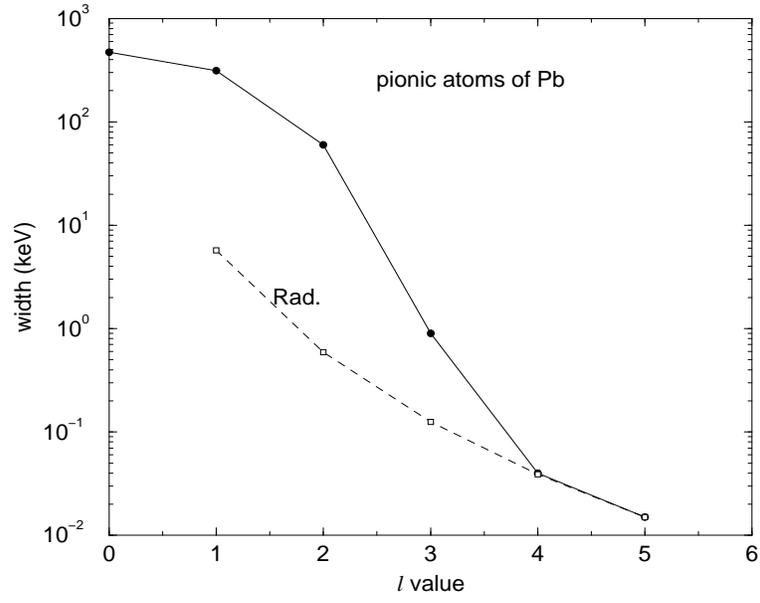}
\caption{Radiation (dashed) and total widths of pionic atom levels
in Pb.}
\label{fig1}
\end{figure}

The motivation for presenting this overview is the revival 
in recent years of interest in
the pion-nucleus interaction at threshold  and in
particular in its relation to the corresponding pion-nucleon interaction.
For  low energy pions one may relate the two
interactions by the Ericson Ericson (EE) potential~\cite{EEr66},
which is inserted into the Klein-Gordon (KG) equation \cite{BFG97}
to calculate shifts and widths of pionic atom levels.  Traditionally,
experiments on pionic atoms involved the measurement of X-rays from
atomic transitions of $\pi ^-$ which terminate when the nuclear
absorption of pions becomes dominant. 
It was believed that in any case the deepest atomic levels 
for heavy nuclei will be too
broad to be well-defined. Friedman and Soff \cite{FSo85} predicted in 1985
that such states, owing to the repulsive $s$-wave $\pi N$ interaction
at threshold,
 are sufficiently narrow so as to be well defined, and 
three years later Toki and Yamazaki discussed ways to populate such
states \cite{TYa88}. The first observation of these states was by
Yamazaki et al.~\cite {YHI96}. Figure \ref{fig1} illustrates how 
nuclear absorption becomes dominant 
when the angular momentum~$l$ goes down and how its initial
increase is greatly reduced due to the repulsion of the wavefunction out
of the nucleus, which is the mechanism responsible for making the
level widths sufficiently small to be observed.

\begin{figure}
\includegraphics[height=11cm,width=10cm]{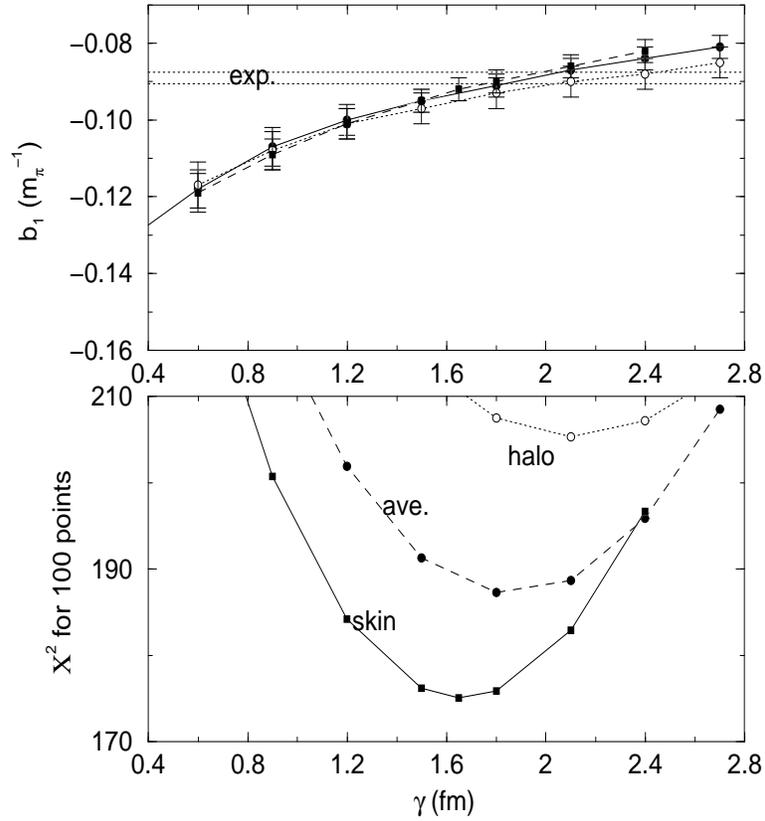}
\caption{$\chi ^2$ {\it vs.} the neutron-excess radius parameter 
$\gamma$ for different
shapes of the neutron density (lower part); derived values of the
isovector parameter $b_1$ (upper).}
\label{fig2}
\end{figure}

Interest  in the pion-nucleus interaction
at low energies
has been focused on the $s$-wave part of the 
optical potential \cite{BFG97}

\begin{eqnarray} \label{equ:EE1s}
q(r) & = & -4\pi(1+\frac{\mu}{M})\{{\bar b_0}(r)
[\rho_n(r)+\rho_p(r)]
  +b_1[\rho_n(r)-\rho_p(r)] \} \nonumber \\
 & &  -4\pi(1+\frac{\mu}{2M})4B_0\rho_n(r) \rho_p(r) ,
\end{eqnarray}
where $M$ is the mass of the nucleon and 
$\rho_n$ and $\rho_p$ are the neutron and proton density
distributions normalized to the number of neutrons $N$ and number
of protons $Z$, respectively. The function ${\bar b_0}(r)$
is given in terms of the {\it local} Fermi
momentum $k_{\rm F}(r)$ corresponding to the isoscalar nucleon
density distribution:
\begin{equation} \label{equ:b0b}
{\bar b_0}(r) = b_0 - \frac{3}{2\pi}(b_0^2+2b_1^2)k_{\rm F}(r).
\end{equation}
This term shows enhanced repulsion compared to the free
pion-nucleon interaction, resulting mostly
from the  in-medium
isovector $s$-wave $\pi N$ amplitude $b_1$
which plays a dominant role due to the nearly vanishing of the corresponding
isoscalar amplitude $b_0$. Some extra repulsion comes also from the empirical
dispersive component of the two-nucleon absorption term $B_0$.
Recent interest in this topic included attempts to explain the
enhancement of the $s$-wave repulsion
 in terms of chiral-motivated density dependence of the pion decay constant
\cite{Wei01} or by energy-dependent effects~\cite{ETa82,FGa03}.

In Section \ref{sec:dens} we discuss the 
role of the neutron density distributions
which, together with the proton distributions, form an essential ingredient
of the EE potential \cite{EEr66,BFG97}. In Section \ref{sec:atoms} we present
results from global fits to pionic atom data, with emphasis placed on
dependence on the neutron distribution used in the analysis. Recent
extensions to the scattering regime are presented in Section \ref{sec:scatt}.
Section \ref{sec:sum} is a summary.

\section{The Role of Neutron Densities}
\label{sec:dens}

With nine parameters in the EE pion-nucleus optical potential \cite{BFG97}
the only
way of gaining significant information from fits to pionic atom 
data is to perform large scale fits to as wide a data base as possible.
The proton density distributions $\rho _p$ are known quite well from
electron scattering and muonic X-ray experiments, and can be obtained 
from the nuclear charge distributions by numerical unfolding of the 
finite size of the proton. In contrast, the neutron densities $\rho _n$
are not known to sufficient accuracy and we have therefore performed fits
while scanning over these densities. This procedure is based on the expectation
that for a large data set over the whole of the periodic table some local
variations will cancel out and that an {\it average} behavior may 
be established.

Experience showed \cite{FGa03} that
the feature of  neutron density distributions which is most relevant
in  determining strong interaction effects
in pionic atoms is the radial extent, as represented for
example by $r_n$, the neutron density rms radius. Other features
such as  the detailed shape of the distribution have only minor effect.
For that reason we chose the rms radius as the prime parameter in the
present study and adopted the two-parameter Fermi distribution both for 
the proton and for the neutron density distributions as follows:

\begin{equation}
\label{equ:2pF}
\rho_{n,p}(r)  = \frac{\rho_{0n,0p}}{1+{\rm exp}((r-R_{n,p})/a_{n,p})}~~ .
\end{equation}
The use of single-particle densities is not expected to be more 
appropriate when there are many nuclei far removed from closed shells and
when in any case parameters of $\rho _n$ are being varied.
In order to allow for possible differences in the shape of the neutron
distribution, the `skin' and the `halo' forms of Ref. 9 were
used, as well as an average of the two. 
In this
way we have used three shapes of the neutron distribution for each value
of its rms radius all along the periodic table.
The rms radius $r_n$ for the various nuclei was parameterised using the
following expression for the neutron-proton differences

\begin{equation} \label{equ:RMF}
r_n-r_p = \gamma \frac{N-Z}{A} + \delta \; ,
\end{equation}
and scanning over the values of the parameter $\gamma$. This approach 
has been made in analyses of antiprotonic atoms data \cite{TJL01,FGM05}.

\section{Results for Pionic Atoms}
\label{sec:atoms}

\begin{figure}
\includegraphics[height=11cm,width=10cm]{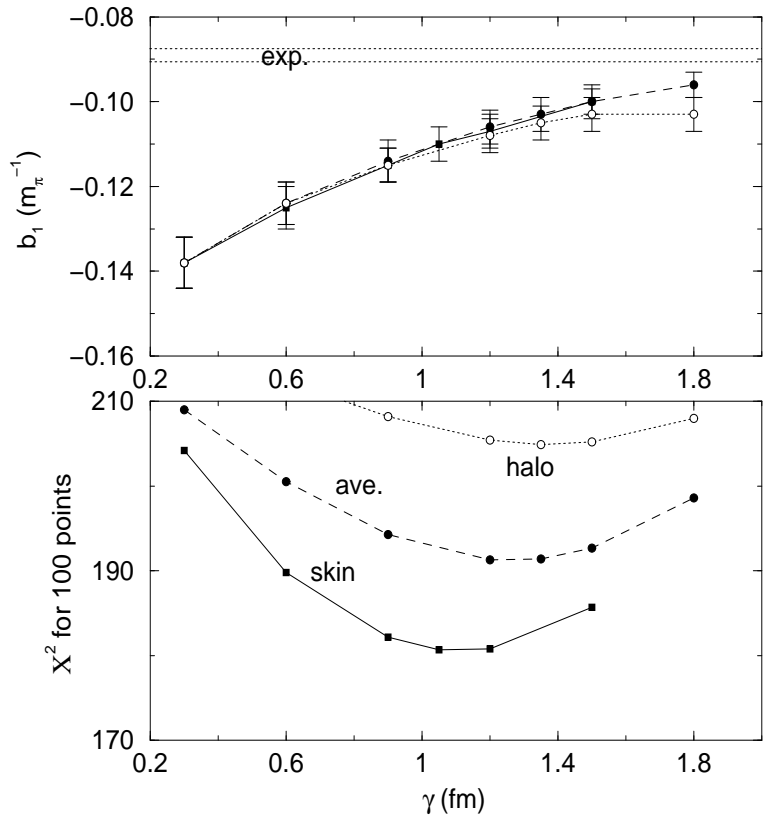}
\caption{Same as Fig. \ref{fig2} but for finite range in the $p$-wave 
term.}
\label{fig3}
\end{figure}

Figure \ref{fig2} shows values of $\chi ^2$ for 100 data points
in the lower part and the derived values of the $s$-wave parameter
 $b_1$ in the upper part, as functions of the 
neutron radius parameter
$\gamma$. It is seen that the best fit is obtained for the `skin' shape
and that the  $b_1$ parameter is then in good agreement with its 
free-space value. We note, however, that this best fit 
is obtained for
a value of $\gamma$ which leads to unacceptably large neutron rms radii
in heavy nuclei \cite{Jas04}. Attempting to introduce finite-range folding
into the potential, as was successfully done with 
antiprotons \cite{FGM05}, the fits deteriorate monotonically with increasing
range. However, when finite range is introduced separately into the
$s$-wave and the $p$-wave parts of the potential, it is found that a
finite range with an rms radius of 0.9$\pm$0.1 fm 
applied only to the $p$-wave part,
leads to the best fit and 
Fig. \ref{fig3} shows that it is obtained for 
$\gamma$ close to 1.0~fm, which is acceptable when comparing with 
other sources of information on neutron radii. The discrepancy between the value
obtained for $b_1$ and its free-space value is clearly observed.
It is also found that Re$B_0$/Im$B_0$=$-$2, which is unacceptable \cite{BFG97}.
Both results represent 
the well known `anomaly', or enhanced repulsion in nuclei.

\begin{figure}
\includegraphics[height=11cm,width=10cm]{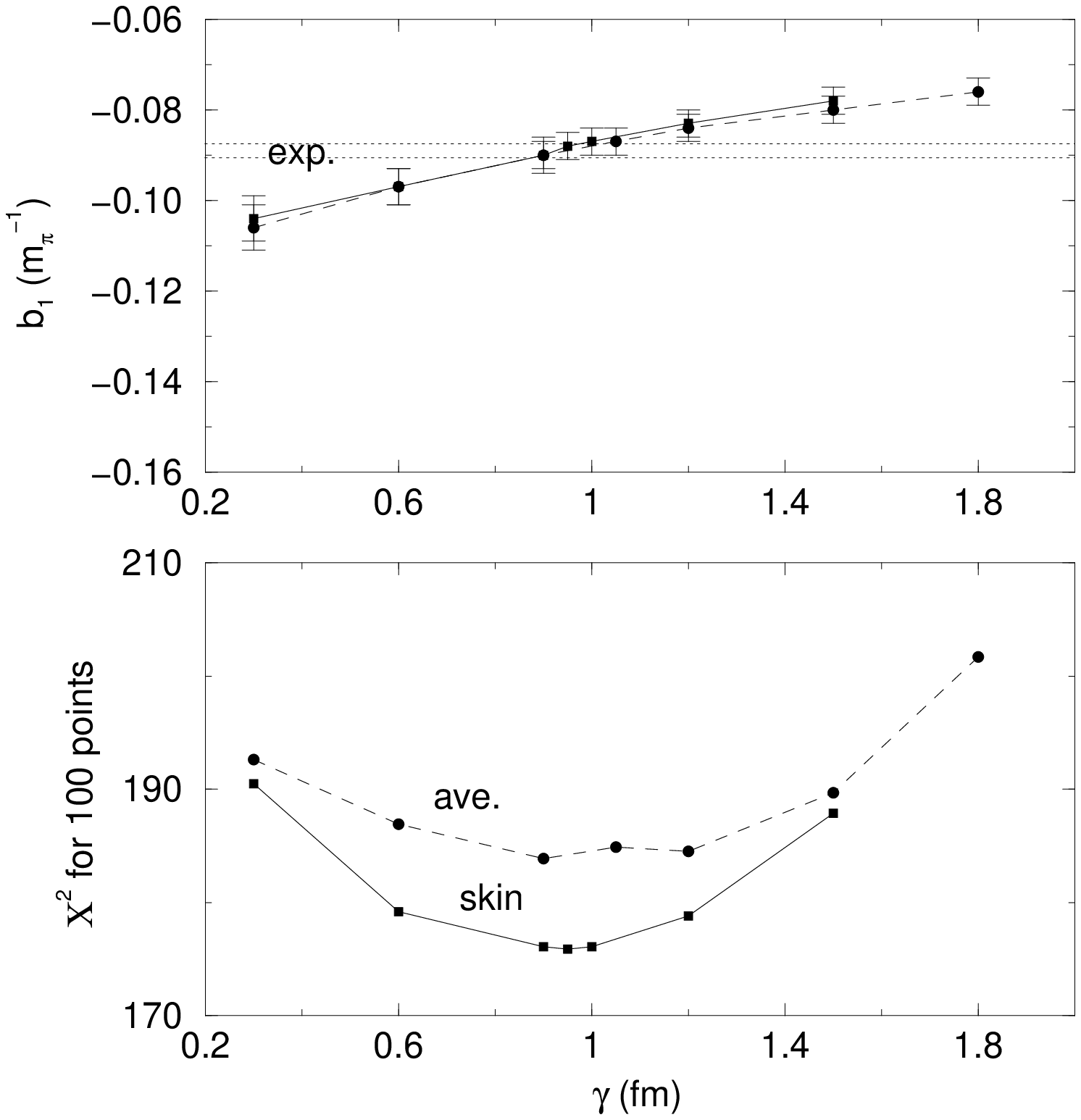}
\caption{Same as Fig. \ref{fig3} but with the chiral model and energy
dependence applied to the $s$-wave 
term.}
\label{fig4}
\end{figure}

In the chiral-approach of Weise \cite{Wei01} 
the in-medium $b_{1}$ is related to
possible partial restoration of chiral symmetry in dense matter,
as follows.
Since $b_{1}$ in free-space is well approximated in lowest
chiral-expansion order by the Tomozawa-Weinberg expression
$b_{1}=-\mu_{\pi N}/8 \pi f^{2}_{\pi}=-0.08~m^{-1}_{\pi}$,
then it can be argued that $b_{1}$ will be modified in pionic atoms
if the pion decay constant
$f_\pi$ is modified in the medium. The square of this decay constant
is given in leading order,
 as a linear function of the nuclear density,
\begin{equation} \label{eq:fpi2}
f_\pi ^{*2} = f_\pi ^2 - \frac{\sigma }{m_\pi ^2} \rho
\end{equation}
with $\sigma$ the pion-nucleon sigma term.
This leads to a density-dependent isovector amplitude such that $b_1$ becomes
\begin{equation}\label{eq:ddb1}
b_1(\rho) = \frac{b_1(0)}{1-2.3\rho}
\end{equation}
for $\sigma $=50 MeV 
and with $\rho$ in units of fm$^{-3}$. With this {\it ansatz} the best
fit is obtained for $\gamma$ close to 1.0 fm and with $b_1$ almost in
agreement with the free value. When the energy dependence of the $b_0$
parameter is also included \cite{ETa82,FGa03}, Fig. \ref{fig4} shows
a perfect agreement between the derived $b_1$ and its free value, for
acceptable neutron rms radii. For this case the dispersive
term Re$B_0$ (not shown)
turns out to be consistent with zero, in contrast to the
unacceptably large repulsive values obtained for the conventional 
potential. It is worth mentioning that although the values of $r_n$
obtained here agree with the values found from analyses of antiprotonic
atoms \cite{TJL01,FGM05}, the latter favor the `halo' shape for the neutron
density distribution. This could be the result of using the over-simplified
Fermi distribution and the fact that whereas pionic atoms are sensitive to 
densities around 50\% of the central value, antiprotonic atoms sample
regions where the density is only 5\% of that of nuclear matter.

Finally, it should be noted that the `deeply bound' pionic atom states
fit precisely into the picture emerging from global fits to conventional
pionic atom data in the sense that predictions made with potentials obtained
from fits to the latter are in full agreement with experimental results
for the former. That is a natural consequence of the repulsion of the
atomic wavefunction out of the nucleus which prevent really deep
penetration of the deeply-bound atomic wavefunctions.

\section{Elastic Scattering at 21.5 MeV}
\label{sec:scatt}

With the picture emerging from global analyses of pionic atoms
it is interesting to extend the study of the $s$-wave part of the
pion-nucleus potential into the scattering regime, where below 30 MeV
the pions penetrate well into nuclei. 
In the scattering scenario, unlike in the atomic
case, one can study both charge states of the pion, thus increasing
sensitivities to isovector effects and to the energy dependence
of the isoscalar amplitude due to the Coulomb interaction. Note that
both were found to play a role in the atomic case.
For that reason  precision measurements of elastic scattering
of 21.5 MeV $\pi ^+$ and $\pi ^-$ by several nuclei were performed very
recently at PSI \cite{FBB04,FBB05} and analyzed in terms of the 
same effects as 
in pionic atoms. 
The experiment was dedicated to studying the elastic scattering of both
pion charge states and special emphasis was placed on the absolute
normalization of the cross sections, which was based on the parallel
measurements of Coulomb scattering of muons.
The potentials tested were the conventional (C) one,
the chiral motivated potential of Weise \cite{Wei01} (W), the energy-dependent
potential \cite{ETa82,FGa03} (E) and a potential with both effects
included (EW).

\begin{table}
\caption{Values of $b_1$ from fits to elastic scattering
of 21.5 MeV $\pi ^\pm$ by Si, Ca, Ni and Zr. $b_1=-0.088\pm0.001 
(m_\pi ^{-1}$) for the free $\pi N$ interaction. (See Ref. 12,13)}
\label{tab:scatt}
\begin{tabular}{ccccc} \hline
model &  C & W & E & EW \\
$b_1 (m_\pi ^{-1}$)& $-0.114\pm0.006$\hphantom{00} &
 \hphantom{0}$-0.081\pm0.005$ & \hphantom{0}$-0.119\pm0.006$ &$-0.083\pm0.005$\\
$\chi ^2$ for 72 points & 134 & 88 & 80 & 88 \\ \botrule
\end{tabular}
\end{table}

Table \ref{tab:scatt} summarizes the results, showing that clearly the better
fits to the data require that at least one of these effects is included, and
that the derived $b_1$ agrees with the free $\pi N$ interaction only
when the chiral-motivated density dependence is included.

\section{Summary}
\label{sec:sum}
Global analyses of strong interaction effects in pionic atoms, fitting
parameters to the EE potential across the periodic table, consistently
led to `anomalous' repulsion in the $s$-wave part of the potential
when compared to the free $\pi N$ interaction at threshold. This is
particularly clear when the dependence on rms radii of neutron
density distributions is considered. Introducing into the $s$-wave
part of the potential  a chiral-motivated dependence on density
of the isovector interaction and
the energy dependence of the isoscalar interaction fully explain the
enhanced repulsion and the best fit is then obtained with neutron
densities that are in agreement with other sources of information.
Dedicated experiment at 21.5 MeV measuring the elastic scattering of 
$\pi ^{\pm}$ by several targets show that the pion-nucleus potential
changes smoothly from threshold into the scattering regime. Enhanced
repulsion is observed also at 21.5 MeV, and  is accounted for
by the same mechanisms as for pionic atoms. However, 
in contrast with pionic atoms, the quality
of fits to the data  clearly require the inclusion of at least
the chiral-motivated density dependence.

\section*{Acknowledgments}
I wish to thank A. Gal for many years of collaboration on topics
of the present manuscript. This work was supported in part
by the Israel Science Foundation grant 757/05.

\end{document}